\documentclass[aps,prb,twocolumn,showpac,spreprintnumbers,amsmath,amssymb,superscriptaddress]{revtex4}

\usepackage{graphicx}
\usepackage{dcolumn}
\usepackage{bm}
\usepackage{color}
\usepackage[colorlinks, allcolors = blue]{hyperref}
\usepackage{lipsum}

\usepackage{ulem}
\usepackage{float}

\newcommand{\ket}[1]{|#1\rangle}

\newcommand{\be}{\begin{equation}}
\newcommand{\ee}{\end{equation}}
\newcommand{\bea}{\begin{eqnarray}}
\newcommand{\eea}{\end{eqnarray}}

\newcommand{\fig}[1]{Fig.~\ref{#1}}

\newcommand{\up}{\uparrow}
\newcommand{\down}{\downarrow}

\newcommand{\ie}{i.\,e. }

\begin{document}

\title{Cross-correlations in a quantum dot Cooper
pair splitter with ferromagnetic leads}

\author{Piotr Trocha}
\email{ptrocha@amu.edu.pl}\affiliation{Faculty of Physics, Adam
Mickiewicz University, 61-614 Pozna\'n, Poland}

\author{Kacper Wrze\'sniewski}
\affiliation{Faculty of Physics, Adam
Mickiewicz University, 61-614 Pozna\'n, Poland}

\date{\today}

\begin{abstract}
We investigate Andreev transport through a quantum dot attached to
two external ferromagnetic leads and one superconducting electrode.
The transport properties of the system are studied by means of the
real-time diagrammatic technique in the sequential tunneling regime.
To distinguish various contributions to Andreev current we calculate the current cross-correlations,
i.e., correlations between currents flowing through two junctions with normal leads.
We analyze dependence of current cross-correlations on various parameters of the considered
model, both in linear and nonlinear transport regimes. The processes and mechanisms leading to enhancement,
suppression or sign change of current cross-correlations are examined and discussed.
Interestingly, our results show that  for specific transport regimes
splitted Cooper pair results in two uncorrelated electrons.
However, utilizing ferromagnetic leads instead of non-magnetic
electrodes can result in positive current cross-correlations.

\end{abstract}



\maketitle


\section{Introduction}\label{Sec:Introd}

Electronic transport phenomena in hybrid quantum dot (QD) systems have recently attracted
great attention \cite{rodero11,konig09,konigPRB10,
eldridgePRB10,csonka,sadov,wojcikPRB14,weymannPRB14,trochaPRB14,trochaPRB15,feinberg,melin16}.
Particularly, QD systems with one superconducting lead and two normal metal
electrodes enable creation of nonlocal entangled electron pairs \cite{sunSC01,sauret,chenAPL04,feng06,golubev07}. Moreover, splitting of Cooper pairs into
two spatially separated electrodes has been demonstrated experimentally in a carbon nanotube double quantum dot
system \cite{hofstetterSC,herrmann10SC,hofstetterPRL11}.
These investigations are important both from the fundamental point of view and
also for future applications in quantum computing \cite{deFranceschiNat11}.

When the applied bias voltage window is in the superconducting gap, the current flows mainly due to
Andreev reflection processes \cite{andreev} while the quasiparticle tunneling becomes negligible in
the low temperature limit. Generally, in-gap tunneling processes can occur via direct Andreev reflection (DAR),
crossed Andreev reflection (CAR), or elastic cotunneling (EC).
Under certain conditions, by properly changing device parameters,
one can tune the contributions due to CAR and
DAR, or EC processes or even suppress one of them. Thus, a tool which allows
to distinguish these contributions is sought. An important one seems to be given by current cross-correlations,
i.e., correlations calculated  between currents flowing through two junctions with normal leads \cite{ChevallierPRB11,rechPRB12}.

The current cross-correlations can deliver a
deeper insight into tunneling processes contributing
to electronic transport \cite{blanter} and
has been reported experimentally \cite{suk,mc,das}
in various setups. Generally, positive current cross-correlations
can be attributed with interactions supporting currents in both
junctions. Especially, they are present in systems with
superconducting electrodes \cite{martin,torres,lesovik,bignon,melin,dong,freyn,freynPRL,riwar,wrzesien},
like Cooper pair splitters, in which enhancement of
positive current cross-correlations can be associated with
high Cooper pair splitting efficiency.
On the other hand, positive current cross-correlations
can be suppressed by interactions, which mutually block the currents
flowing through two junctions. The tunneling processes that occur
in opposite directions are associated with negative sign of current
cross-correlations.

Interestingly, negative current cross-correlations in electron beam splitter device  have been observed experimentally in a Hanbury Brown-Twiss setup \cite{henny,oliver}. Moreover, in a paramagnetic multiterminal quantum dot current cross-correlations have been found to be negative
\cite{bagrets}. However, in a three-terminal QD system coupled to ferromagnetic leads some positive cross-correlations between output currents can appear as a result of dynamical spin blockade on the dot  not observed in the paramagnetic case \cite{cottet}.

The current cross-correlations are also useful for
determining the efficiency and fidelity of an electron-entangler device and
allow to formulate Bell-type inequalities \cite{chtch,sauret2}. Their violation
provides an evidence of nonlocality of split pairs of electrons.

One should also bear in mind that positive current cross-correlations
noticed in hybrid superconducting structures are not always
due to CAR processes \cite{melin,floser}. For instance, such situation
has been predicted for normal metal-superconductor-normal metal
hybrid structures with highly transparent interfaces \cite{floser}.

In hybrid QD's systems being in proximity to superconductor the formation of entangled electrons' pairs seems to be rather natural. Therefore, superconducting lead acts as a source of entangled pairs of electrons as its ground state is occupied by Cooper pairs in a spin-singlet state. Furthermore, these Cooper pairs can be extracted from a superconductor by tunneling through the dot's states into the normal metal leads. However, two Andreev tunneling processes have to be distinguished: direct Andreev reflection (DAR) and crossed Andreev reflection (CAR). The former process occurs when Cooper pair extracted from superconductor is transmitted to the same normal lead, whereas in the latter one, entangled pair of electrons leaving the superconductor is split  into its individual electrons which end in two spatially-separated leads. A minimal condition to assure the split electrons stay correlated is that the coherence length of Cooper pair has to exceed the width of the superconducting source contact.  In an efficient electron entangler, which converts a charge current to a flux of spin-entangled electron pairs, the CAR processes have to be enforced. By analyzing current cross correlations one is able to distinguish regions in device's parameters space with high and low Cooper pairs splitting efficiency. Particularly, large positive current cross-correlations can be associated with enhancement in Cooper pair splitting efficiency, while small values of the aforementioned quantity indicate low splitter's efficiency.

In this paper we study
dependence of current cross-correlations on various parameters of the considered
model focusing on the in-gap transport regime.
The processes and mechanisms leading to enhancement or suppression
of current cross-correlations are examined and discussed. Moreover, the influence of
external magnetic field and ferromagnetism of external leads on current cross-correlations
 is also investigated. We use the
real-time diagrammatic technique to calculate transport characteristics.
In general, the external weakly coupled leads are assumed to
be metallic ferromagnetic electrodes. We consider collinear
magnetic configurations, \ie when magnetic moments of both leads
are aligned in the same directions (parallel) or oppositely (antiparallel).

The paper is organized in the following way:
In Section \ref{Sec:theory} we introduce the model of the system and present theoretical background.
This includes description of the model and also description
of the method used to calculate quantities of interest.
In Section \ref{Sec:num} numerical results are presented and discussed.
In this section we distinguish the
cases of leads with and without spin polarization. Additionally,
the cases of system with and without external
magnetic field are considered separately.
Finally, Section \ref{Sec:conc} includes brief summary and general conclusions.


\section{Theoretical description}\label{Sec:theory}

\begin{center}
\begin{figure}[h]
\includegraphics[width=0.9\columnwidth]{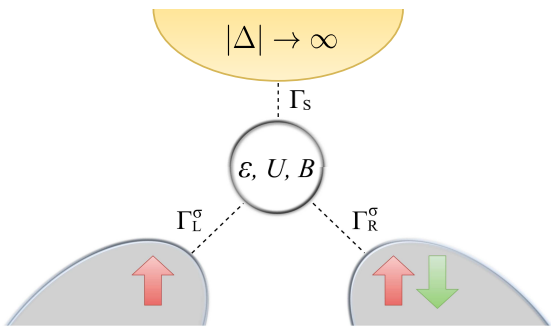}
\caption{\label{Fig:1} Schematic picture of the QD system
  coupled to two normal metal and one superconducting leads.}
\end{figure}
\end{center}

The system taken into considerations consists of single-level quantum dot attached
to two normal metal and one superconducting lead as shown in Fig.~\ref{Fig:1}.
The Hamiltonian of the system acquires the following form:
\begin{equation}\label{Eq:1}
  H=\sum_{\beta=L,R}H_{\beta} + H_S +H_{QD}+H_T,
\end{equation}
where the first term, $H_{\beta}$ for $\beta=L,R$, describes the left ($L$) and
right ($R$) electrodes in the noninteracting quasiparticle
approximation, $ H_{\beta}=\sum_{\textbf{k}\sigma}\varepsilon_{\textbf{k}\beta\sigma}
c_{\textbf{k}\beta\sigma}^{\dagger}c_{\textbf{k}\beta\sigma}$ with $\varepsilon_{\textbf{k}\beta\sigma}$ denoting the single particle
energy.

The second term in Eq.~(\ref{Eq:1}) describes the s-wave BCS superconducting lead in the mean field approximation
\begin{eqnarray}\label{Eq:2}
H_{S}&=&\sum_{{\mathbf
k},\sigma}\limits\varepsilon_{{\mathbf k}S}
     c^\dag_{{\mathbf k}S}c_{{\mathbf k}S}
     \nonumber \\
     &+&
     \sum_{{\mathbf k}}\limits\left(\Delta^{\ast} c_{{\mathbf
k}S\downarrow}c_{{-\mathbf k}S\uparrow}+\Delta c^{\dag}_{{-\mathbf
k}S\uparrow}c^{\dag}_{{\mathbf k}S\downarrow}\right),
\end{eqnarray}
with $\varepsilon_{{\mathbf k}S}$ denoting the relevant
single-particle energy and  $\Delta$, assumed real and positive, standing
for the order parameter of the superconductor.

The third term of Hamiltonian describes single-level quantum dot and acquires the following form:

\begin{equation}\label{Eq:3}
  H_{QD}=\sum_{\sigma}\varepsilon_{\sigma}d_{\sigma}^{\dagger}d_{\sigma}+Un_{\uparrow}n_{\downarrow},
\end{equation}
where $\varepsilon_{\sigma}$ and $U$  denote the spin dependent dot's energy level and the corresponding
Coulomb integral. Applying external magnetic field dot's energy level becomes split, $\varepsilon_{\sigma}=
\varepsilon\pm B/2$ with $B$ denoting Zeeman splitting energy.

Finally, tunneling of electrons between the leads ($L,R,S$) and the quantum dot can be modelled by the Hamiltonian
\begin{equation}\label{Eq:4}
H_T=\sum_{\mathbf{k}\sigma}\limits\sum_{\beta=L,R,S}
   \limits (V_{\mathbf{k}\sigma}^\beta c^\dag_{\mathbf{k}\beta\sigma}d_{\sigma}+\rm
   h.c.)
\end{equation}
with $V_{\mathbf{k}\sigma}^\beta$ denoting the relevant tunneling matrix elements.

In the wide band approximation dot's coupling to the normal metal electrodes can
be assumed to be energy independent and constant, $\Gamma_{L}^{\sigma}=\Gamma_L(1+\tilde{\sigma}p)$, and
$\Gamma_{R}^{\sigma}=\Gamma_R(1+\eta\tilde{\sigma}p)$ with $\tilde{\sigma}=1$ for $\sigma=\up$
 and $\tilde{\sigma}=-1$ for $\sigma=\down$. Here, $p$ denotes the spin polarization
 of magnetic leads assumed the same for the left and right electrodes, whereas
 $\eta=\pm 1$ for parallel (upper sign) and antiparallel (lower sign)
 magnetic alignment of ferromagnetic leads. Furthermore, we assume symmetric couplings,
$\Gamma_L=\Gamma_R\equiv\Gamma/2$.

As we are interested in Andreev transport regime we can take the limit of an infinite superconducting gap, $\Delta\rightarrow\infty$. Then the quantum dot coupled to the superconducting lead is described by an effective Hamiltonian \cite{rozhkov}:
\begin{equation}\label{Eq:5}
  H_{QD}^{\rm eff}=\sum_{\sigma}\varepsilon_{\sigma}d_{\sigma}^{\dagger}d_{\sigma}+Un_{\uparrow}n_{\downarrow}-
  \frac{\Gamma_S}{2}\left(d_{\uparrow}^{\dagger}d_{\downarrow}^{\dagger}+d_{\downarrow}d_{\uparrow}\right)
\end{equation}
where the effective pair potential $\Gamma_S$ is the coupling strength between the dot and superconducting electrode and acquires the form $\Gamma_S=2\pi|V^{S}|^2\rho_{S}$ with $\rho_{S}$ denoting BCS density of states in the normal state.

%
The eigenstates of the effective dot's Hamiltonian acquire the form: $|\sigma\rangle$, and
$|\pm\rangle=1/\sqrt{2}\left(\sqrt{1\mp\delta/(2\varepsilon_A)}|0\rangle\mp\sqrt{1\pm\delta/(2\varepsilon_A)}|2\rangle\right)$, while
the corresponding eigen-energies are: $E_\uparrow=\varepsilon_\uparrow$, $E_\downarrow=\varepsilon_\downarrow$, and $E_{\pm}=\delta/2\pm\varepsilon_A$, with $\delta=\varepsilon_\uparrow+\varepsilon_\downarrow+U$. Here, $\varepsilon_A=\sqrt{\delta^2+\Gamma_S^2}/2$ measures
the energy difference between the states $|+\rangle$ and $|-\rangle$.

The Andreev bound states' energies are defined as:
\begin{equation}\label{Eq:6}
E_{\alpha,\beta}^A=\alpha\frac{U}{2}+\frac{\beta}{2}\sqrt{\delta^2+\Gamma_S^2},
\end{equation}
where $\alpha,\beta=\pm$. These energies are the excitation
energies of the dot decoupled from the normal metal leads.

In order to determine the transport properties of the system we employ the
real-time diagrammatic technique \cite{schoeler, thielmann, governalePRB08}. The stationary occupation probability $p^{st}_{\chi}$
of a state $\ket{\chi}$ can be found from:
\begin{equation}\label{Eq:master}
   \mathbf{W} \mathbf{ p^{st}}=0 ,
\end{equation}
where $\mathbf{ p^{st}}$ is the vector containing probabilities $p^{st}_{\chi}$
and the elements $W_{\chi\chi'}$ of self-energy matrix $\mathbf{W}$
account for transitions between the states $\ket{\chi}$ and $\ket{\chi'}$.

The current flowing through the junction with lead $\alpha=(L,R)$
can be found from:
\begin{equation}\label{Eq:8}
  I_{\alpha}=\frac{e}{2\hbar} {\rm Tr} \left\{ {\mathbf W}^{I_\alpha}{\mathbf p}^{st} \right\},
\end{equation}
where the self-energy matrix ${\mathbf W}^{I_\alpha}$
is similar to ${\mathbf W}$, but it takes into account the number
of electrons transferred through a given junction.

The current cross-correlations in the sequential tunneling approximation, $S_{LR}$, are defined as~\cite{thielmann}:
\begin{equation}\label{Eq:Slr}
  S_{LR}=\frac{e^2}{\hbar} {\rm Tr} \left\{ \left[
  {\mathbf W}^{I_L}\mathbf{P} {\mathbf W}^{I_R}+
  {\mathbf W}^{I_R}\mathbf{P} {\mathbf W}^{I_L}\right]
  {\mathbf p}^{st} \right\},
\end{equation}
where the propagator ${\mathbf P}$ is determined
from $\mathbf{WP} = \mathbf{p}^{st}\mathbf{e}^T - \mathbf{1}$,
with $\mathbf{e}^T = (1,1,\dots,1)$.

\section{Numerical results}\label{Sec:num}
We present the numerical results for Andreev transport assuming large
superconducting-gap limit. Here, we show the differential conductance
associated with current injected/extracted into/from superconductor
and the corresponding current cross-correlations.
The differential conductance, $G_S$, is calculated as
$G_S=dI_S/dV$ with $I_S$ denoting Andreev current.
The Andreev current is simply obtained from Kirchoff's law as $I_S=I_L+I_R$.

The section is divided
into two parts. In the first part of the section the case of
nonmagnetic external electrodes ($p=0$) is considered, whereas in the second
part we investigate the case of magnetic leads ($p\neq 0$). Moreover, throughout
the sections we consider two distinct cases, with and without external magnetic
field leading to finite or no Zeeman splitting.

\subsection{Nonmagnetic leads ($p=0$)}\label{Sec:p0}
In this section the transport properties of QD coupled
to one superconducting electrode and two nonmagnetic metallic
leads are considered. The differential conductance and corresponding
current cross-correlations are calculated for two
cases. The former case corresponds to situation when no external
magnetic field is applied, whereas in the latter one the influence of
magnetic field is taken into account.

\subsubsection{No external magnetic field, $B=0$}\label{Sec:p0B0}
\begin{figure}
 \includegraphics[width=0.9\columnwidth]{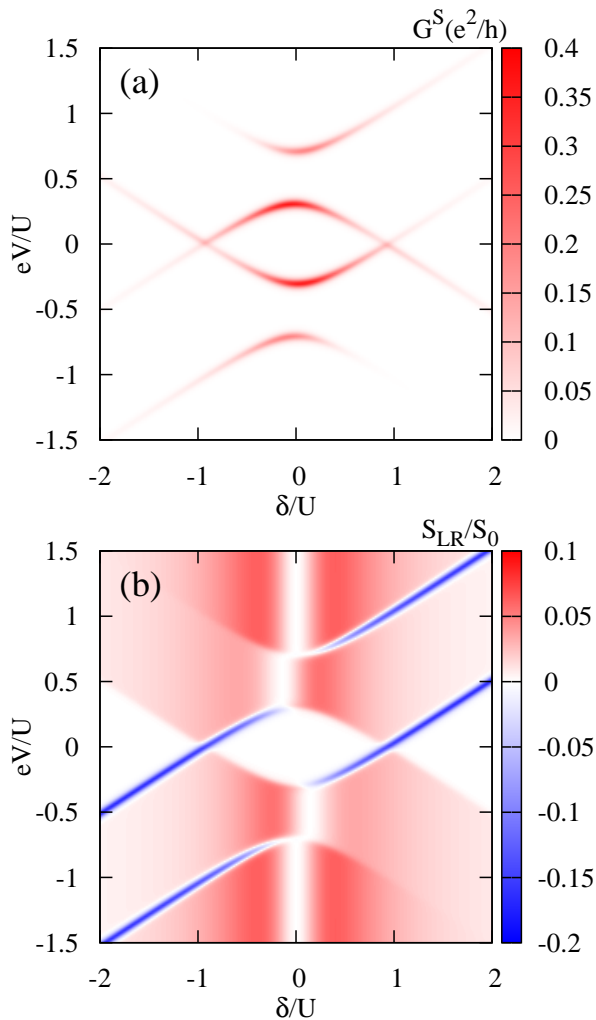}
  \caption{ \label{Fig:2}
  (color online)  Differential conductance $G_S$ (a) and corresponding
  current cross-correlations (b)
  as a function of bias voltage and detuning $\delta$ calculated for
  magnetic field, $B=0$.
  The other parameters are: $U=1$ (used as energy unit), $\Gamma_S=0.4$,
  $\Gamma=0.01$ and $T=0.015$, with $S_0 = e^2\Gamma/\hbar$.}
\end{figure}

The differential conductance $G_S$ corresponding to Andreev current
and the respective current cross-correlations $S_{LR}$ calculated for zero magnetic field, $B=0$,
are shown in \fig{Fig:2} as a function of bias voltage
applied to the two normal leads. The system is biased in the following way:
$\mu_L=\mu_R=eV$ and grounding the superconducting electrode $\mu_S=0$.
We introduce detuning parameter, $\delta = 2 \varepsilon + U$, for spin-degenerate dot's
level, \ie for $\varepsilon_\up=\varepsilon_\down\equiv\varepsilon$.

First of all, one can notice that differential conductance obeys the symmetry, $G_S(\delta, eV)=G_S(-\delta,-eV)$.
Due to the fact that the Andreev current (not shown) is optimized when particle-hole symmetry holds,
it becomes significant only for small detuning $\delta$. As a result, $G_S$ acquires the largest values
for $\delta$ close to zero. The differential conductance, $G_S$, reveals a peak each time the electrochemical
potential of normal metal leads crosses one of the
Andreev levels. Generally, for zero magnetic field, $G_S$ exhibits four peaks associated with four Andreev
levels. However, for $|\delta|=\sqrt{U^2-\Gamma_S^2}$ only three peaks appear as the states $E_A^{+-}$
and $E_A^{-+}$ become degenerate. Moreover, $G_S$ is asymmetric with respect to the bias reversal for finite detuning
$\delta\neq 0$, which is directly related with the behavior of the Andreev current.

In Fig.~\ref{Fig:2}(b) the corresponding current cross-correlations, $S_{LR}$, are shown.
One can notice that $S_{LR}$ vanishes in the Coulomb blockade regime when the dot is occupied by
single electron, i.e., for $E_{+-}<eV<E_{-+}$ and for $|\delta|<\sqrt{U^2-\Gamma_S^2}$.
This is the result of vanishing current in the Coulomb blockade regime as two electrons are
required to form Cooper pair.
The current cross-correlations also vanish for $|\delta|>\sqrt{U^2-\Gamma_S^2}$ and for small bias voltage,
$E_A^{+-}<eV<E_A^{-+}$, when no current flows. These regions corresponds
to empty or doubly occupied dot regime.
For most regions, where the current can flow, $S_{LR}$ acquires positive values, which
indicates that CAR processes give main contribution to the current.
Interestingly, $S_{LR}$ exhibits also negative values. Specifically, $S_{LR}$
becomes negative when bias voltage passes through $E_{-+}^A$ for $\delta \gtrsim 0$
and for $\delta\lesssim 0$ when $eV\approx E_{+-}^A$. Another two regions of negative current
cross-correlations appear for $\delta >0$ when $eV\approx E_{++}^A$ and for
$\delta <0$ when $eV\approx E_{--}^A$.
This happens each time the bias voltage $eV$ approaches dot's level $\varepsilon$ or $\varepsilon+U$.
Then, single electron can tunnel from the left lead onto the dot and re-tunnel into the right lead. Reverse tunneling processes
($R\rightarrow L$) occur with the same probability. Hence, there is no net charge current, but these processes give
contribution to the current cross-correlation, specifically to its negative values, indicating that tunneling
processes by left and right junctions occur in opposite directions and compensate each other.

It is worth noting that $S_{LR}$ vanishes in the vicinity of particle-hole symmetry point
($\delta\approx 0$) for $eV>E_{++}^A$ and for $eV<E_{--}^A$, irrespective of the presence of magnetic field.
However, it can not be stated that Andreev current flows
only due to DAR processes although $S_{LR}$ vanishes. According to symmetry of the system, both
DAR and CAR processes equally contribute to the Andreev transport.
It is enough to consider positive bias case with $eV>E_{++}^A$ as for negative bias voltage with $eV<E_{--}^A$
an analogous discussion can be applied.
Due to strong coupling between the dot and superconducting lead ($\Gamma_S\gg\Gamma$) fast coherent oscillations
of Cooper pairs occur between $QD$ and superconductor. These coherent oscillations are occasionally interrupted by
tunneling of single electron from normal lead (left or right). Average interval of time between tunneling of
an electron from normal lead into the dot is relatively large ($2\hbar/\Gamma$) comparing with the Cooper pairs
oscillations' period ($2\hbar/\Gamma_S$). Thus, from the point of view of normal leads, single electrons tunnel
into the dot independently (of the oscillations) \cite{BraggioSSC}. Thus, such uncorrelated single-electron tunneling events
of two subsequent electrons originating from different normal leads can not give impact to $S_{LR}$.
To support the above explanation let us present more formal considerations.
The zero-frequency cross correlations between the currents flowing through the left and right junctions
can be defined as follows:
\begin{equation}
    S_{LR} = \int^{\infty}_{-\infty} dt \langle \delta I_L(t) \delta I_R(0) + \delta I_R(0) \delta I_L(t) \rangle,
\end{equation}
with $\delta I_{\alpha}(t) = \hat{I}_{\alpha}(t) - \langle \hat{I}_{\alpha} \rangle$ and $\hat{I}_{\alpha}$ being the current operator.

For the bias voltages $eV>E_{++}^A$ and for $eV<E_{--}^A$ in the particle-hole symmetry point ($\delta/U=0$) the Andreev current is maximized, but simultaneously the current cross-correlations are completely suppressed, \ie $S_{LR}=0$. This behavior can be explained by analysing the states' probabilities of the quantum dot and matrix elements of current matrices $W^{\mathrm{I_\alpha}}$ for discussed transport parameters.
The probability of occupation of each available states, \ie $|+\rangle, |-\rangle, |\uparrow\rangle$ and $|\downarrow\rangle$ is the same and equals to $p_\chi=1/4$ with $\chi=+,-,\uparrow,\downarrow$.
Moreover, all non-zero matrix $W^{\mathrm{I_\alpha}}$ elements for both left and right junctions are equal as well, which means
that all possible tunneling events have equal and maximal rates, with left/right junction indifference.
This results in a constant average current in each (left/right) junction of equal value.
Moreover, it turns out that the currents through both junctions are constant in time.
This has been checked by solving rate equation $ \dot{\mathbf{p}}(t) =  \mathbf{W} \mathbf{p}(t)$
and calculating time evolution of currents through left and right junction.
As a result, there are no fluctuations of the current away from the average value, \ie
$\delta I_{\alpha}(t) = \hat{I}_{\alpha}(t) - \langle \hat{I}_{\alpha} \rangle = 0$, and thus, $S_{LR}=0$.

Similar considerations can be applied for bias voltage $E_{-+}^A<eV<E_{++}^A$ as well as for $E_{--}^A<eV<E_{+-}^A$. However, now the zero
current cross-correlations are realized for slightly shifted $\delta$ from particle-hole symmetry point. In the former case, $S_{LR}=0$
is noticed around $\delta=-\frac{\sqrt{2}}{4}\Gamma_S$, whereas for the latter one, current cross-correlations vanish for $\delta=\frac{\sqrt{2}}{4}\Gamma_S$.

\subsubsection{Finite external magnetic field, $B\neq0$}\label{Sec:p0Bnon0}
\begin{figure}
 \includegraphics[width=0.9\columnwidth]{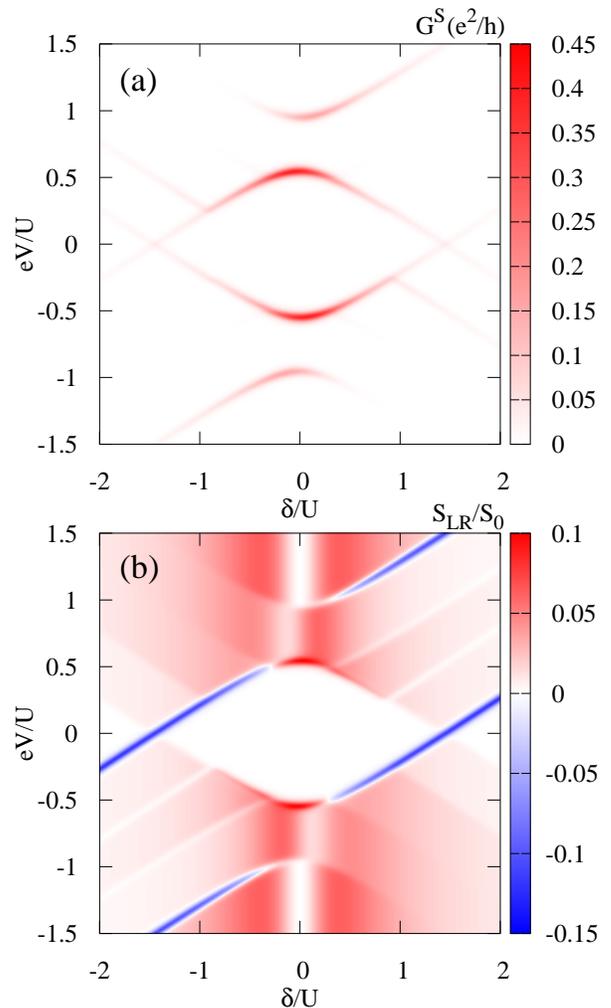}
  \caption{ \label{Fig:3}
  (color online) Differential conductance $G_S$ (a) and corresponding
  current cross-correlations (b)
  as a function of bias voltage and detuning $\delta$ calculated for
  magnetic field, $B/U=0.5$.
  The other parameters are as in \fig{Fig:2}.}
\end{figure}

Applying finite magnetic field leads to the splitting of the Andreev bound states and to doubling of
corresponding excitation energies. Now, Andreev excitation energies acquire the following form,
$E_{\alpha,\beta,\gamma}^A=E_{\alpha,\beta}^A+\gamma \frac{B}{2}$ with $\gamma=\pm 1$.
As a result, more peaks appear in the differential conductance, see e.g. Fig.~\ref{Fig:3}(a)
calculated for $B/U=0.5$. One can notice that the range of Coulomb blockade becomes enlarged,
as filling the dot with second electron cost the energy $\varepsilon_\uparrow +U$.

In the case of finite magnetic field the current cross-correlations reveal rather minor
qualitative and quantitative differences disregarding the effect due to Andreev bound states splitting.
First of all, the maximal positive values of current cross-correlations
are of the same order as these corresponding to zero magnetic field case.
In turn, the negative values of current
cross-correlations become slightly increased.
Due to additional Andreev bound states originating from Zeeman splitting $S_{LR}$
exhibits more steps. However, the number of regions where $S_{LR}$ is negative becomes independent of magnetic field.
In the remaining regions between the steps $S_{LR}$ vanishes instead of acquiring negative values.
The most striking qualitative difference in $S_{LR}$
generated by magnetic field is associated an enhancement of positive current cross-correlations when $eV$
crosses through $E_{+-+}^A$ ($E_{-+-}^A$) for $\delta\in (-\sqrt{B^2-\Gamma_S^2},\sqrt{U^2-\Gamma_S^2})$
 ($\delta\in (-\sqrt{U^2-\Gamma_S^2},\sqrt{B^2-\Gamma_S^2},)$), \ie for bias voltages for which
 QD passes from singly occupied state to doubly or empty state, respectively, allowing for
 CAR processes to occur.
The widths of these peaks in $S_{LR}$ becomes smaller with decreasing temperature and disappear
in the limit $T\rightarrow 0$.

\subsection{Magnetic leads ($p>0$)}\label{Sec:pnon0}
So far, the calculations have been performed for QD
Cooper pair beam splitters with nonmagnetic leads.
However, because using ferromagnetic leads can be important to estimate
entanglement between split electrons~\cite{klobus}, here, we provide comprehensive
study of transport properties of the QD Cooper beam splitters with ferromagnetic contacts.

In this section, the effects arising due to magnetism of external weakly coupled leads
are considered. Now, the left and right electrodes are
spin polarized and the strength of the
external leads' polarization is described by spin polarization
factor, $p$, assumed equal for both ferromagnetic electrodes.
Here, we consider collinear magnetic configurations of
ferromagnetic leads: parallel (${\mathrm P}$) and antiparallel (AP).
The former one corresponds to the situation when magnetic moments
of ferromagnetic leads are aligned in the same direction,
whereas in the latter one they align in the opposite directions.

\begin{figure}[t]
 \includegraphics[width=0.9\columnwidth]{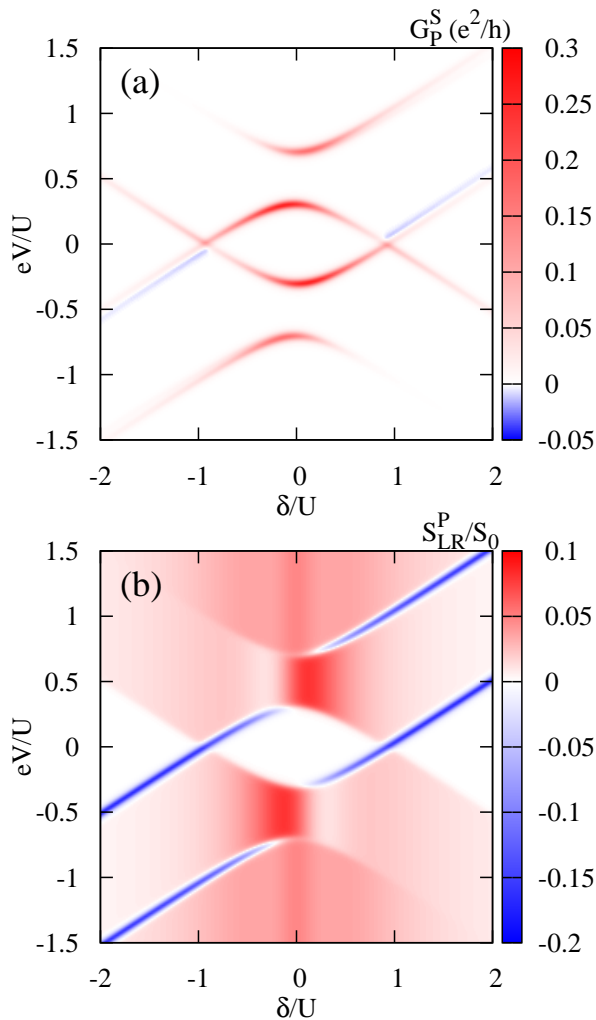}
  \caption{ \label{Fig:4}
  (color online)  Differential conductance $G_S$ (a) and corresponding
  current cross-correlations (b)
  as a function of bias voltage and detuning $\delta$ in
  parallel magnetic configuration P calculated for spin polarization $p=0.5$.
  The other parameters are as in Fig.~\ref{Fig:2}.}
\end{figure}

Let us first consider parallel magnetic configuration. In this case the
differential conductance [\fig{Fig:4}(a)] reveals several differences comparing
with nonmagnetic situation [\fig{Fig:2}(a)]. One can notice that
maximal intensity of differential conductance becomes suppressed
as the minority carriers determine the Andreev current. Accordingly,
in the parallel magnetic configuration the density
of states of minority carriers rules the rate of injecting/extracting
electron pairs, which becomes the {\it bottleneck}
for the Andreev transport.
However, more striking
difference is appearance of negative values in the differential
conductance. Specifically, negative values of the differential
conductance emerge for $|\delta|>\sqrt{U^2-\Gamma_S^2}$ in the
vicinity of Andreev levels $E_{-+}^A$ ($E_{+-}^A$)
for positive (negative) electrochemical potential shift, see
Fig. 4(a). The physical mechanism leading to negative differential conductance
in hybrid three-terminal QD system
is due to the nonequilibrium spin accumulation
in the dot and has been explained in details in Ref.~\onlinecite{weymannPRB14}.

\begin{figure}[t]
 \includegraphics[width=0.9\columnwidth]{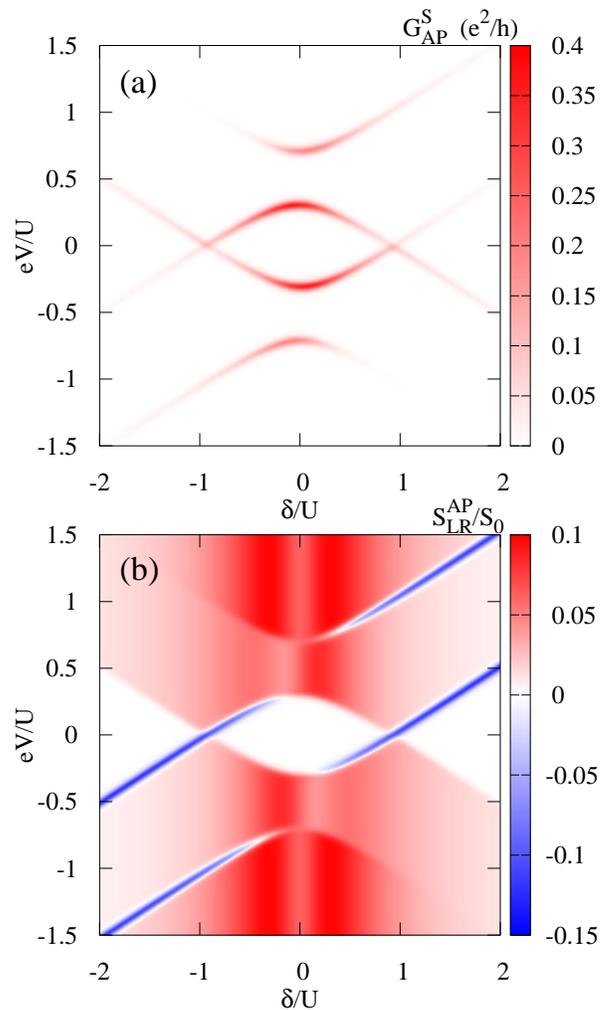}
  \caption{ \label{Fig:5}
  (color online)  Differential conductance $G_S$ (a) and corresponding
  current cross-correlations (b)
  as a function of bias voltage and detuning $\delta$ in
  antiparallel magnetic configuration AP calculated for spin polarization $p=0.5$.
  The other parameters are as in Fig.~\ref{Fig:2}.}
\end{figure}

In turn, in the antiparallel magnetic configuration (\fig{Fig:5}) the
differential conductance $G_S^{AP}$  behaves similarly to that one obtained for $p=0$
because total
population of electrons with a given spin direction
in both leads remains constant disregarding of the value
of $p$.

On the other hand, current cross-correlations exhibit both qualitative
and quantitative differences in both magnetic configurations
comparing with $p=0$ case. First of all, the current cross-correlations
become enhanced in the vicinity of particle-hole symmetry point
for $eV>E_{++}^A$ and for $eV<E_{--}^A$ in both magnetic configurations
(compare Figs.\ref{Fig:4}, \ref{Fig:5} with \fig{Fig:2}).

However, maximal positive values
of $S_{LR}$ in the ${\mathrm P}$-alignment become suppressed, whereas negative
values are slightly increased (in the sense of absolute values)
which can be clearly seen in \fig{Fig:4}.
In turn, $S_{LR}$ in AP configuration
differs significantly from that in ${\mathrm P}$ configuration
and also from that in the nonmagnetic case. One can notice
that current cross-correlations for antiparallel magnetic configuration
reach larger maximal values comparing with $S_{LR}$ for both
parallel alignment and $p=0$ case.

To provide a deeper insight into $p$ dependence of Andreev current ($I_S$) and
corresponding current cross-correlation $S_{LR}$, in both magnetic configurations,
we derive
some approximative analytical formulas assuming low temperature limit.
Here, we concentrate on the particle-hole symmetric case, \ie{$\delta\approx 0$}
as for $\delta\neq 0$ the obtained formulas become cumbersome.
The corresponding formulas for $\delta\approx 0$ and for
$\frac{1}{2}[U-\Gamma_S]\lesssim eV\lesssim\frac{1}{2}[U+\Gamma_S]$
are:
\bea\label{P1}
I_S^P&=&\frac{4(1-p^2)}{3+p^2}I_0
\nonumber \\
S_{LR}^P&=&\frac{1+21p^2-29p^4+7p^6}{2(3+p^2)^3}S_0
\eea
for parallel ($P$),
and
\bea\label{AP1}
I_S^{AP}&=&\frac{4}{3}I_0
\nonumber \\
S_{LR}^{AP}&=&\frac{1+9p^2}{54}S_0
\eea
for antiparallel (AP) alignment, where $I_0=e\Gamma/\hbar$.
The analytical formulas for $S_{LR}$ at negative bias voltage $eV<0$ in
the corresponding regions are the same as for $eV>0$, whereas
that for current are obtained from the relation, $I_S(\delta,eV)=-I_S(-\delta,-eV)$.

The above formulas confirm the deduced $p$-dependence of current-current cross correlations.
Specifically, $S_{LR}$ in parallel magnetic configuration changes nonmonotonically with
increasing $p$. For small values of polarization factor $S_{LR}$ grows
with increasing $p$ until achieving maximum at $p=\sqrt{1/23(27-8\sqrt{6})}\approx 0.57$.
It is worth noting that such a value of spin polarization is observed for typical ferromagnets.
With further increase of $p$, $S_{LR}$
decreases with increasing $p$. Finally,
for half-metallic leads ($p=1$) current-current cross-correlations vanish and so does the current.
In turn, current cross-correlations in AP alignment monotonically grow with
increasing $p$ achieving maximum for half-metallic leads ($p=1$).

The next step in the current, for $\delta\approx 0$, appears at $eV\gtrsim\frac{1}{2}[U+\Gamma_S]$. In this regime, in the parallel
magnetic configuration, the current and the corresponding current cross-correlations are described by the following formulas:
\bea\label{P2}
I_S^P&=&2(1-p^2)I_0
\nonumber \\
S_{LR}^P&=&\frac{p^4(1-p^2)}{4}S_0.
\eea
The above formula shows that $S_{LR}$ is non-negative for all values of polarization
factor and nonmonotonically depends on $p$ achieving maximum at $p=1/\sqrt{2}$.
In AP alignment current becomes independent from spin polarization factor $p$ and reaches its maximal value $I_S^{AP}=2I_0$,
while for the current-current cross-correlations one finds $S_{LR}^{AP}=p^2/4$. Increase of $S_{LR}^{AP}$
is a result of enhanced CAR processes at the expense of DAR processes, which become suppressed while increasing $p$.

Moreover, an applied magnetic field influences current and corresponding
current cross-correlations. At the first step in the current
one finds the following formulas for $I_S$ and $S_{LR}$ calculated for $B/U=0.5$:
\begin{widetext}
\bea\label{BP1}
I_S^P&=&\frac{(p-1)(p+1)(p+3)}{2}I_0
\nonumber \\
S_{LR}^P&=&\frac{(1-p)(1+p)(2-10p+5p^2+12p^3+3p^4)}{64}S_0.
\eea
\end{widetext}

in P configuration, and
\bea\label{BAP1}
I_S^{AP}&=&\frac{3}{2}I_0
\nonumber \\
S_{LR}^{AP}&=&\frac{2+13p^2}{64}S_0,
\eea
in ${\mathrm AP}$ configuration.
Accordingly, for bias voltages at which the second step in the current occurs both $I_S$ and $S_{LR}$
are not influenced by magnetic field and resulting formulas remain the same as in
the absence of magnetic field.

\begin{figure}
 \includegraphics[width=0.9\columnwidth]{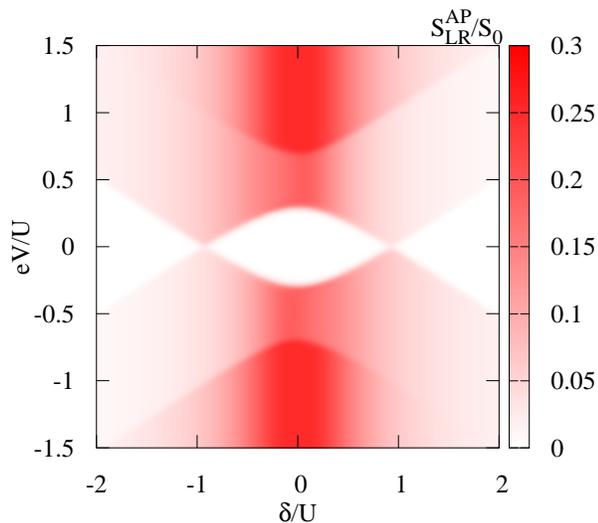}
  \caption{ \label{Fig:6}
  (color online) Current cross-correlations
  as a function of bias voltage and detuning $\delta$ in
  antiparallel magnetic configuration $AP$ calculated for spin polarization $p=1$.
  The other parameters are as in Fig.~\ref{Fig:2}.}
\end{figure}

Finally, in the case of half-metallic leads in AP alignment
current-current cross-correlations acquire non-negative values in
the whole range of bias voltage and detuning parameter $\delta$
(see \fig{Fig:6}).
This is a result of total blockage of tunneling processes through
the left and right junctions in opposite directions, as there are no
available states in a given ferromagnetic lead for electrons incoming
to the other ferromagnetic electrode. Moreover, both current and
current-current cross-correlations become maximized.
Maximization of current cross-correlations is well understood
as for half-metallic leads in the AP alignment only CAR
processes can contribute to Andreev current, whereas DAR processes
are totally blocked, since in given electrode only electrons with
one spin orientation are available. From the experimental
point of view, such a device can be used for verification
of the presence and role of CAR processes.
Although, ideal half-metallic leads are rather unaccessible
in real systems, one can utilize the change in magnetic configuration
of magnetic electrodes from antiparallel to parallel alignment
in order to quantify the role of CAR comparing to DAR.
Such protocol can be used because any switch of magnetic alignment
of the system affects only CAR, whereas DAR processes
are not influenced. By comparing the transport properties of the
system in both magnetic configurations, some information on CAR
can be extracted.

\section{Conclusions}\label{Sec:conc}

In this paper we have analyzed the current cross-correlations and differential conductance
corresponding to the Andreev current
flowing in single quantum dot based Cooper pair splitters.
We have shown that finite magnetic field leads to the splitting of Andreev bound states, which
can be clearly seen in the differential conductance characteristics. The current cross-correlations
exhibit both positive and negative values in zero and finite magnetic field.
However, finite magnetic field tends to suppress negative values (in the sense of absolute value)
of the current cross-correlations and
simultaneously it enhances the positive values of this quantity at least in specified regions of transport parameters.
Moreover, we have found that in the vicinity of particle-hole symmetry point the current cross-correlations
are suppressed to zero and the splitting pairs of electrons ending in two separate normal metal leads become uncorrelated.
However, using ferromagnetic leads, in place of nonmagnetic ones, can restore entanglement of pair of split electrons in this
transport regime. This finding is important as one wants to obtain large current of entanglement split pairs of electrons and as the Andreev current achieves maximum values in the particle-hole symmetry point.

When external weakly coupled leads are ferromagnetic the behavior of transport properties
become more interesting. Especially, we have found negative differential conductance in the
parallel magnetic configuration. Moreover, significantly different behavior of current
cross-correlations has been found in both magnetic configurations compared to the nonmagnetic case.
Furthermore, the dependence of current cross-correlations on leads' polarization factor $p$ has been addressed
for both magnetic alignments. Specifically, in the parallel magnetic configuration the
current-current cross-correlations exhibit nonmonotonic behavior with varying $p$; they reach
maximal value for finite polarization factor suggesting that competition
between CAR and DAR processes in P alignment is not trivial. On the other hand, in
the antiparallel magnetic configuration contribution of CAR processes gradually grows
with increasing $p$, and finally for half-metallic leads $p=1$, CAR becomes the only source
of Andreev current.
We have also indicated possible experimental utilization of the considered system
in order to quantify the different processes contributed to the transport.

\begin{acknowledgements}
K.W. acknowledges support from the Polish National Science Centre from funds awarded
through the decision No. DEC-2013/10/E/ST3/00213. The authors thank I. Weymann for
discussions.
\end{acknowledgements}

\end{document}